\documentclass[a4paper,11pt]{article}
\usepackage{pos}
\usepackage{physics}

\newcommand{\unit}[1]{\;\mathrm{#1}}
\newcommand{\me}[0]{~\mathrm{e}}
\renewcommand{\vec}[1]{~\mathbf{#1}}

\title{Formation of light (anti)nuclei}

\author*[a]{J. Tjemsland}

\affiliation[a]{Institutt for fysikk, NTNU,\\
  Trondheim, Norway}

\emailAdd{jonas.tjemsland@ntnu.no}

\abstract{The production mechanism of light nuclei, such as
deuteron, helium-3, tritium and their antiparticles, 
has recently attracted an increased attention from the astroparticle
and heavy ion communities. The expected low astrophysical background of light
antinulei makes them ideal probes for exotic astrophysical processes, such as
dark matter annihilations. At the same time, they can be used to measure
two-nucleon correlations and density fluctuations in heavy ion collisions, which
may shed light on the QCD phase diagram. Motivated by the importance of light
antinuclei in cosmic ray studies, we developed a new coalescence model for light
(anti)nuclei that includes both the size of the formation
region, which is process dependent, and momentum correlations
in a semi-classical picture. We have employed the model as an
afterburner to the event generators Pythia 8 and QGSJET II, and find that the
model agrees well with experimental data on antideuteron and antihelium-3
production in $e^+e^-$, $pp$, $p$Be and $p$Al collisions at various energies.
In this paper, we review this model and update existing fits
to experimental data based on new insights.
}

\FullConference{%
  Tools for High Energy Physics and Cosmology - TOOLS2020\\
  2-6 November, 2020\\
  Institut de Physique des 2 Infinis (IP2I), Lyon, France}

\begin{document}
\maketitle

\section{Introduction}

The production mechanism of light nuclei, such as deuteron, helium-3, tritium
and their antiparticles, has attracted attention from both the
astroparticle and heavy ion communities. These particles are of particular
importance in cosmic ray studies due to their low expected
background at small kinetic energies, for a recent review see Ref.
\cite{vonDoetinchem:2020vbj}. This makes them a promising detection channel
for exotic physics, such as dark matter annihilations and decays.
Currently, the measurement of astrophysical
antinuclei is performed by the AMS-02 experiment on-board the International Space
Station, while the balloon-borne GAPS experiment is expected to be launched 
in December 2022 \cite{Saffold:2020pfc,battiston_antimatter_2008}.
Unfortunately, the AMS-02 collaboration has not yet published any results
regarding antinuclei measurements, even though the expected antideuteron 
flux from secondary production and optimistic dark matter models is close
to its estimated sensitivity \cite{vonDoetinchem:2020vbj}.
In heavy ion collisions, (anti)nuclei measurements are of particular interest
due to their small binding energies. This make them sensitive probes for
two-nucleon correlations and density fluctuations that may shed light on
the QCD phase
diagram \cite{Caines:2017vvs}. In order to correctly interpret the results of
both collider and cosmic ray experiments, a precise description of the
production mechanism is important. The focus in this work is on small
interacting systems.

In Ref. \cite{Kachelriess:2019taq}, we developed a new coalescence model for
deuteron, tritium and helium-3 based on the Wigner function representation of
the produced nuclei states (the abbreviation WiFunC, short for Wigner Function
with Correlations, will be used for this model). The model was later refined and
applied in Ref. \cite{Kachelriess:2020uoh} for astrophysical processes, and in
Ref. \cite{Kachelriess:2020amp} for recent collider data. In contrast to existing
production models, the WiFunC model includes both the size of the nucleon
formation region and two-particle correlations in a semi-classical picture. In
this work, the WiFunC model is reviewed.

This paper is structured as follows. In section \ref{sec:coal} the standard
coalescence model in momentum space commonly used for small interacting
systems is discussed. This serves as a motivation for the WiFunC model, which is
discussed in section \ref{sec:wifunc}. In subsection \ref{ssec:fit} we
update and extend the fits to experimental data from Refs.
\cite{Kachelriess:2019taq,Kachelriess:2020uoh} based on new insights. The
numerical implementation of the WiFunC model is outlined in section
\ref{sec:numerical_implementation}. Finally, we summarise in section
\ref{sec:summary}.

\section{The coalescence model in momentum space} 
\label{sec:coal}

The production of light nuclei%
\footnote{Most of the discussions related to coalescence apply
	equally well to particles as to antiparticles, and the prescription
	``anti'' will thus often be neglected.}
is often described
using coalescence models. In this type of models, a nucleus is produced from
nucleons that have
(nearly) completed their formation. Traditionally, the yield of a nucleus
consisting of $Z$ protons and $N$ neutrons is parametrised as
\begin{equation}
	E_A\dv[3]{N_A}{P_A} = \left.B_A\left(E_p\dv[3]{N_p}{p_p}\right)^Z
		\left(E_n\dv[3]{N_n}{p_n}\right)^N\right|_{P_A/A=p_n=p_p},
\end{equation}
where $B_A$ is the so-called coalescence factor and $A=Z+N$ is the nucleon
number. Motivated by the small binding energies of light nuclei,
the coalescence condition is often applied in momentum space for small
interacting
systems. In this case, two nucleons coalesce if their invariant momentum 
difference is less than the phenomenological coalescence momentum $p_0$
\cite{Schwarzschild:1963zz,butler_deuterons_1963,Chardonnet:1997dv}.
In the limit of isotropic nucleon yields, one can show
that $B_A\propto p_0^{3(A-1)}$. This can in turn be improved by considering the
coalescence condition per-event using momentum correlations provided by
a Monte Carlo generator, as first proposed in Refs.
\cite{kadastik_enhanced_2010,Dal_thesis}. Furthermore, it was later noted in Ref.
\cite{ibarra_prospects_2013} that particles have to be close to the primary
interaction to be able to coalesce, which essentially mean
that weak decays should be considered as a separate cluster
\cite{Winkler:2020ltd}. For heavy ion-collisions, an alternative scheme was
developed where the coalescence factor instead scales with the nucleon emission 
volume as $B_A\propto V^{A-1}$ \cite{Csernai:1986qf,Nagle:1996vp}.

\begin{figure}[htbp]
	\centering
	\includegraphics[scale=0.98]{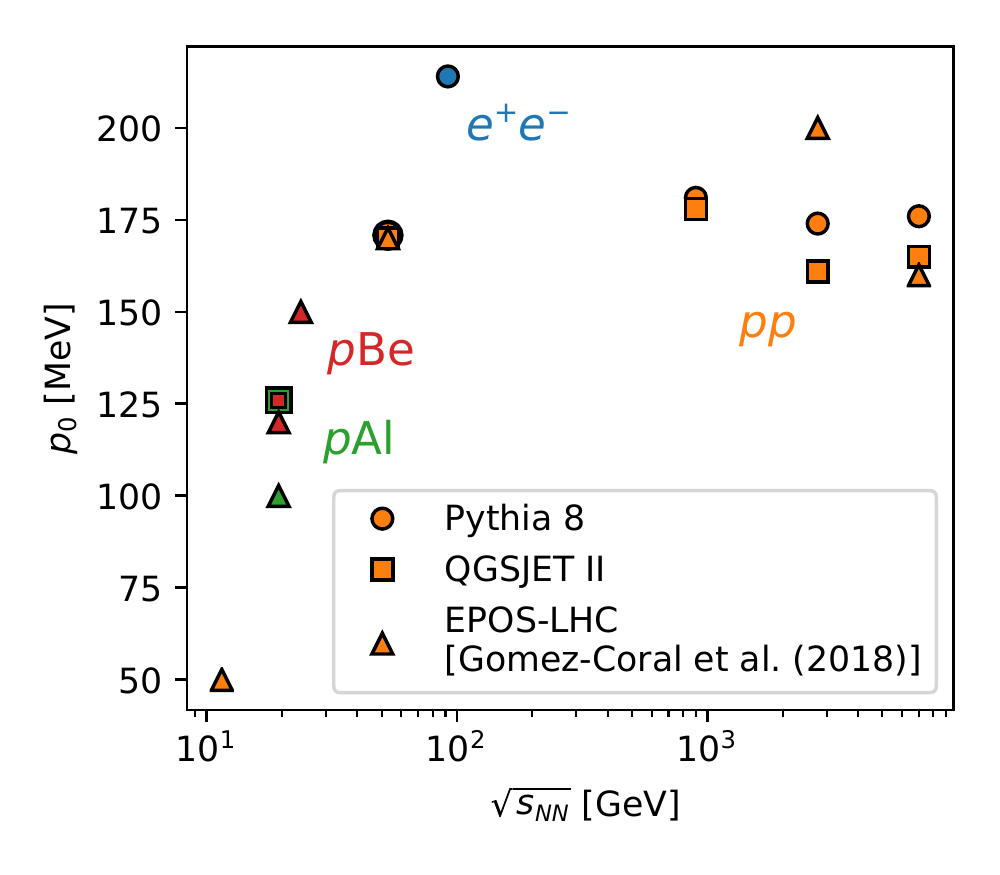}
	\caption{Best fit value of $p_0$ for experiments on $pp$ (orange), $e^+e^-$
	(blue), $pN$ (red and green) interactions using QGSJET II (squares),
	Pythia 8.2 (circles) and EPOS LHC (triangles) from Refs. 
	\cite{Gomez-Coral:2018yuk,Kachelriess:2020uoh}.}
	\label{fig:p0_comparison}
\end{figure}

Clearly, the coalescence model in momentum space is purely phenomenological and
classical. Furthermore, the coalescence momentum $p_0$ should be independent of
both the process type and the center of mass (c.m.) energy to be predictive, 
which is not the case. In Fig. \ref{fig:p0_comparison} we plot the
coalescence momentum $p_0$ obtained by a fit to various
experimental data in Refs. \cite{Gomez-Coral:2018yuk,Kachelriess:2020uoh}. 
As noted in Ref. \cite{Gomez-Coral:2018yuk}, the result in Fig.
\ref{fig:p0_comparison} may suggest that the coalescence momentum $p_0$ exhibits
an energy dependence: $p_0$ rapidly decreases to zero for small $\sqrt{s}$.
There are, however, two problems with this conclusion: The process dependence
has not been accounted for
and the energy dependence has not been explained. These are two motivations for
the WiFunC model, in which the apparent energy dependence is removed by taking
into account the process dependence.\footnote{Note that we have not considered
	the data point at $\sim 12\unit{GeV}$ from Ref. \cite{Abramov:1986ti}. An independent
	confirmation of this data point will	thus falsify our model.}
More importantly, the WiFunC model is semi-classical and has a clear
microphysical interpretation that can be used to predict its free parameter.

\section{The WiFunC model}
\label{sec:wifunc}

\subsection{Motivation}

One of the main goals of the WiFunC model is to take into account both
two-particle correlations and the size of the nucleon formation region when 
describing the production of light nuclei. Two-particle correlations are
naturally important for small interacting systems as the production is highly
non-isotropic. Meanwhile, the importance of the formation region becomes
apparent when inspecting the relevant timescales of the problem
\cite{Dokshitzer:1991wu}: The initial hard scattering occurs on timescales
$t_\mathrm{ann}\sim 1/\sqrt{s}$, and is succeeded by a perturbative cascade with
a characteristic momentum transfer $\Lambda_\mathrm{QCD}^2\ll |q^2|\ll s$. This
means that the largest time and distance scales in the problem is related to the
hadronisation length, $L_\mathrm{had}\sim \gamma L_0$ with $L_0\sim 1\unit{fm}$.
The light nuclei is in turn in the coalescence picture formed by nucleons that
have (nearly) completed their formation. Since the nucleus wave functions have a
size $r^N_\mathrm{rms}\sim \mathrm{few}\unit{fm}$ comparable to the
hadronisation length, the emission volume should not be neglected even in
point-like processes.

\subsection{The underlying physics and theory}
The WiFunC model is based on the quantum mechanical description of coalescence,
as reviewed in e.g. Ref. \cite{scheibl_coalescence_1999}. The deuteron%
\footnote{%
	For concreteness, only the production of deuteron is discussed in this
	subsection.	The model for helium-3 and tritium is derived using the same procedure.}
spectrum can in this picture be found by projecting the reduced density
matrix describing the nucleons onto the nucleus density matrix:
$\dd[3]{N_d}/\dd{P_d}^3=\trace(\rho_d\rho_{nucl})$.
Rewriting this in terms of the two-nucleon Wigner function, $W_{np}$, and the
deuteron Wigner function, $\mathcal{D}$, leads to
\begin{equation}
	\dv[3]{N_d}{P_d}=\frac{3}{8}\int\dd[3]r_d\int\frac{\dd[3]q\dd[3]r}{(2\pi)^6}
		\mathcal{D}(\vec r, \vec q)W_{np}(\vec P_d/2 + q, \vec P_d/2-\vec q,
		\vec r_n, \vec r_p),
	\label{eq:main}
\end{equation}
where the factor $3/8$ is obtained by averaging over spin and isospin.
It is convenient to approximate the deuteron wave function as a Gaussian, in
which case $\mathcal{D}(\vec r, \vec q) =8\exp{-r^2/d^2-q^2d^2}$ with
$d\simeq 3.2\unit{fm}$. The effect of two-particle correlations on the
coalescence probability is, however, sensitive to the shape of the deuteron
wave function. Thus, a more accurate wave function should be used when 
momentum correlations are taken into account, using e.g. the sum of two
Gaussians as introduced Ref. \cite{Kachelriess:2019taq}.\footnote{
	The numerical implementation discussed in section
	\ref{sec:numerical_implementation} includes such an improved wave function.
}

In order to proceed, one has to provide a prescription for the nucleon Wigner
function. In the WiFunC model, a factorisation of the momentum and
position dependence in the Wigner function is assumed,\footnote{
	Note that this implies a transition to a semi-classical treatment.
}
\begin{equation}
	W_{np} = H_{np}(\vec r_n, \vec r_p)G_{np}(\vec P_d/2 + \vec q,
	\vec P_d/2 - \vec q),
	\label{eq:ansatz_1}
\end{equation}
and the spatial nucleon distributions are assumed to be uncorrelated,
$H_{np}(\vec r_n, \vec r_p)=h(\vec r_n)h(\vec r_p)$. In turn, the spatial
distributions are taken to be Gaussians,
\begin{equation}
	h(\vec r) = (2\pi\sigma^2)^{-3/2} \exp{-\frac{r^2}{2\sigma^2}}.
	\label{eq:ansatz_2}
\end{equation}
Inserting the ans\"atze \eqref{eq:ansatz_1} and \eqref{eq:ansatz_2} into Eq.
\eqref{eq:main} leads to the WiFunC model,
\begin{equation}
	\dv[3]{N_d}{P_d} = \frac{3\zeta}{(2\pi)^6} \int\dd[3]{q} e^{-q^2d^2}
	G_{np}(\vec{P}_d/2 + \vec q, \vec{P}_d/2 - \vec q).
	\label{eq:wifunc}
\end{equation}
The function $\zeta$ describes the distribution of the nucleons, and
it is in general given by
\begin{equation}
	\zeta(\sigma_\parallel, \sigma_\perp, d) =
	\sqrt{\frac{d^2}{d^2+4\sigma_\perp^2/(\cos^2\theta + \gamma^2\sin^2\theta)}}
	\sqrt{\frac{d^2}{d^2+4\sigma_\perp^2}}
	\sqrt{\frac{d^2}{d^2+4\sigma_\parallel^2}},
	\label{eq:zeta}
\end{equation}
where $\theta$ is the angle between the deuteron momentum and the $z$-axis
of the c.m. frame of the particle collision, while $\sigma_{\perp/\|}$
describes the
characteristic spatial spread of nucleons in the perpendicular/parallel
direction in the lab frame.\footnote{
	For annihilation processes, $\theta$, $\sigma_\perp$ and
	$\sigma_\|$ is defined relative to the initial quark pair in the hard
	process.
}
The effective Lorentz boost of the transverse spread is included to
account for the boost between the lab frame and deuteron rest frame
\cite{Kachelriess:2019taq}.

In Eq. \eqref{eq:wifunc} one particular choice for the semi-classical emission
volume was used. Some event generators like Pythia 8
\cite{sjostrand_introduction_2015,Ferreres-Sole:2018vgo} and UrQMD
\cite{Bleicher:1999xi} have implemented a semi-classical treatment of the
particle trajectories in the event. In this case, one can instead evaluate
directly
\begin{equation}
	\dv[3]{N_d}{P_d}=3\int\frac{\dd[3]q\dd[3]r}{(2\pi)^6}
		\me^{-r^2/d^2-q^2d^2}W_{np}(\vec P_d/2 + q, \vec P_d/2-\vec q,
		\vec r_n, \vec r_p),
	\label{eq:wifunc_with_r}
\end{equation}
thereby relying on the spatial correlations provided by the event generator
instead of Eqs. \eqref{eq:ansatz_1} and \eqref{eq:ansatz_2}.

\subsection{Process dependence}
\label{ssec:process}

The free parameters in the WiFunC model, $\sigma_\perp$ and $\sigma_\parallel$,
have the physical interpretation as the size of the formation region of
nucleons. They will, in general, have a contribution from the perturbative
cascade and hadronisation, and a contribution related to the finite size of the
colliding particles:
\begin{equation}
	\sigma_{\|,\perp}^2 = \sigma_{\|,\perp (e^+e^-)}^2 +
	\sigma_{\|,\perp (\mathrm{geom})}^2.
	\label{eq:process_dependence_1}
\end{equation}

For point-like processes the spatial spread in
the longitudinal direction is given by $\sigma_\|\simeq R_p\simeq 1\unit{fm}$.
Meanwhile, the spread in the transverse direction comes from the random walk
behaviour of the perturbative cascade. This means that
$\sigma_\perp\simeq \Lambda_\mathrm{QCD}^{-1}\simeq 1\unit{fm}$.

The geometrical contributions can be approximated by
\cite{Kachelriess:2019taq,Kachelriess:2020uoh}
\begin{align}
	\sigma_{\perp\mathrm{(geom)}}^2 &\simeq \frac{2R_1^2R_2^2}{R_1^2+R_2^2},\\
	\sigma_{\|\mathrm{(geom)}}^2 &\simeq \max\left\{R_1,R_2\right\},
\end{align}
where $R_1$ and $R_2$ are the Gaussian radii of the parton clouds in the
colliding particles. For nuclei, these can be well approximated as
\begin{equation}
	R_A\simeq a_0A^{1/3},
	\label{eq:radius_scaling}
\end{equation}
with $a_0\simeq 1\unit{fm}$, where $A$ is the number of nucleons in the
nucleus that take part in the interaction \cite{Donnelly:2017aaa}.
Thus, we can describe the coalescence of
light nuclei using a single universal parameter by setting\footnote{Notice the
ambiguity of this choice: we expect in general
$\sigma_\parallel\gtrsim\sigma_\perp$ and we could therefore have chosen a
different scaling between the parameters. Nevertheless, we find that this
choice well describes the physics and experimental data.
If more accurate data and event generators become available in the future,
a separate fit of $\sigma_\perp$ and $\sigma_\|$, or even the geometrical and
point-like spread, may be needed. }
\begin{equation}
	\sigma\equiv\sigma_{(e^+e^-)}=a_0=\sigma_{(pp)}/\sqrt{2}.
	\label{eq:process_dependence_5}
\end{equation}

\subsection{Comparison with experimental data}
\label{ssec:fit}

\begin{figure}[htbp]
	\centering
	\includegraphics[scale=0.98]{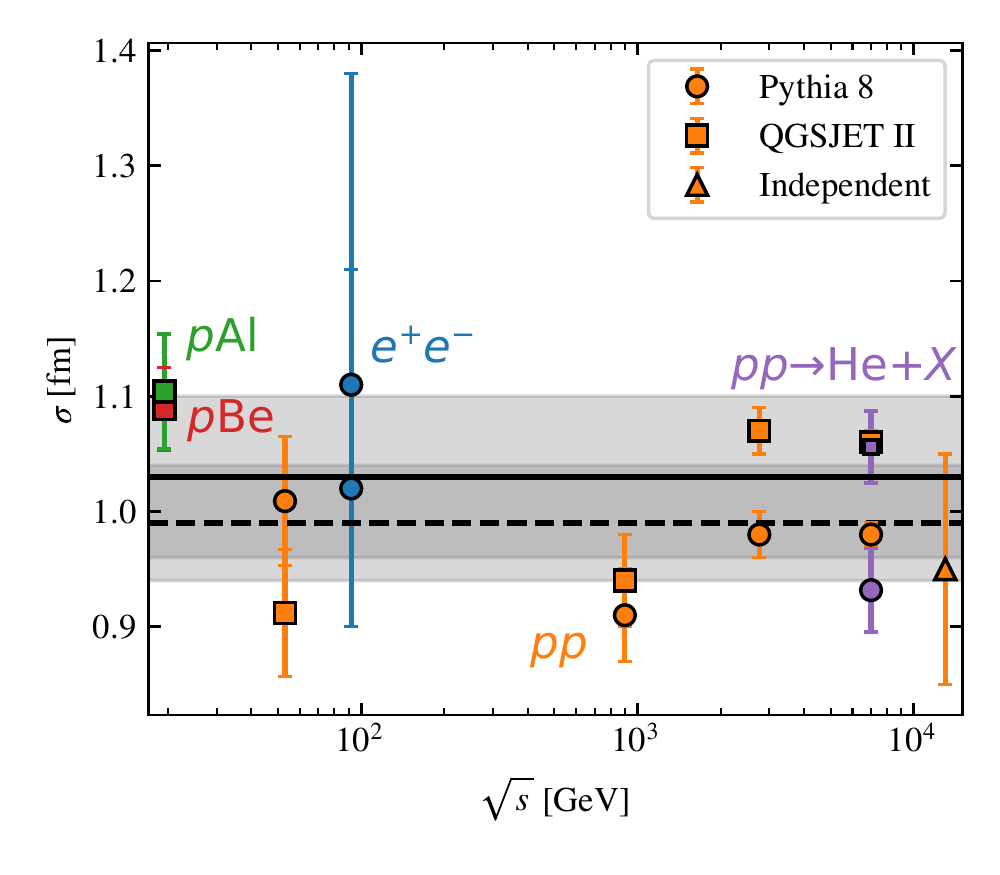}
	\caption{
	The best fit of $\sigma$ to experimental data on antinuclei production
	in $e^+e^-$ (blue), $pp$
	(orange), $p$Be (red) and $p$Al (green) collisions using
	Pythia 8 (circles) and QGSJET II (squares).
	The result obtained in Ref. \cite{Kachelriess:2020amp} from
	comparison with the measured baryon emission volume is also
	shown (triangle).
	The mean of the data points for Pythia (dashed line) and
	QGSJET (solid line) and their standard deviations (transparent gray regions) 
	are plotted without regards to the uncertainties in order to visualise the
	variation.}
	\label{fig:sigma_comparison}
\end{figure}

In Fig. \ref{fig:sigma_comparison} we plot the best fit of $\sigma$ to
the experimental data considered
in Refs. \cite{Kachelriess:2019taq,Kachelriess:2020uoh,Kachelriess:2020amp} 
as a function of the c.m. energy of the collision using Pythia 8
\cite{sjostrand_pythia_2006,sjostrand_introduction_2015}
and QGSJET II \cite{Ostapchenko:2010vb,Ostapchenko:2013pia}.%
\footnote{
In addition to including also the data on helium production for QGSJET, 
we have made the following changes compared to Refs.
\cite{Kachelriess:2019taq,Kachelriess:2020uoh}. First, we now vary the spread
$\sigma$ in $pN$ collisions on an event by event basis depending on the number 
of nucleons that participate in the interaction using Eq.
\eqref{eq:radius_scaling}. Second, we now use the distribution factor $\zeta$
with boost in only one transverse component [Eq. \eqref{eq:zeta}].
Finally, the size of the helium wave function is treated as
discussed in section \ref{ssec:he_wf}.} 
The data points are well described by a constant spread
$\sigma=(1.0\pm0.1)\unit{fm}$, as expected by the WiFunC model.
Thus, the apparent energy dependence seen in Fig. \ref{fig:p0_comparison} is
alleviated by taking into account the nucleon emission volume.

\begin{figure}[htbp]
	\centering
	\includegraphics[scale=0.98]{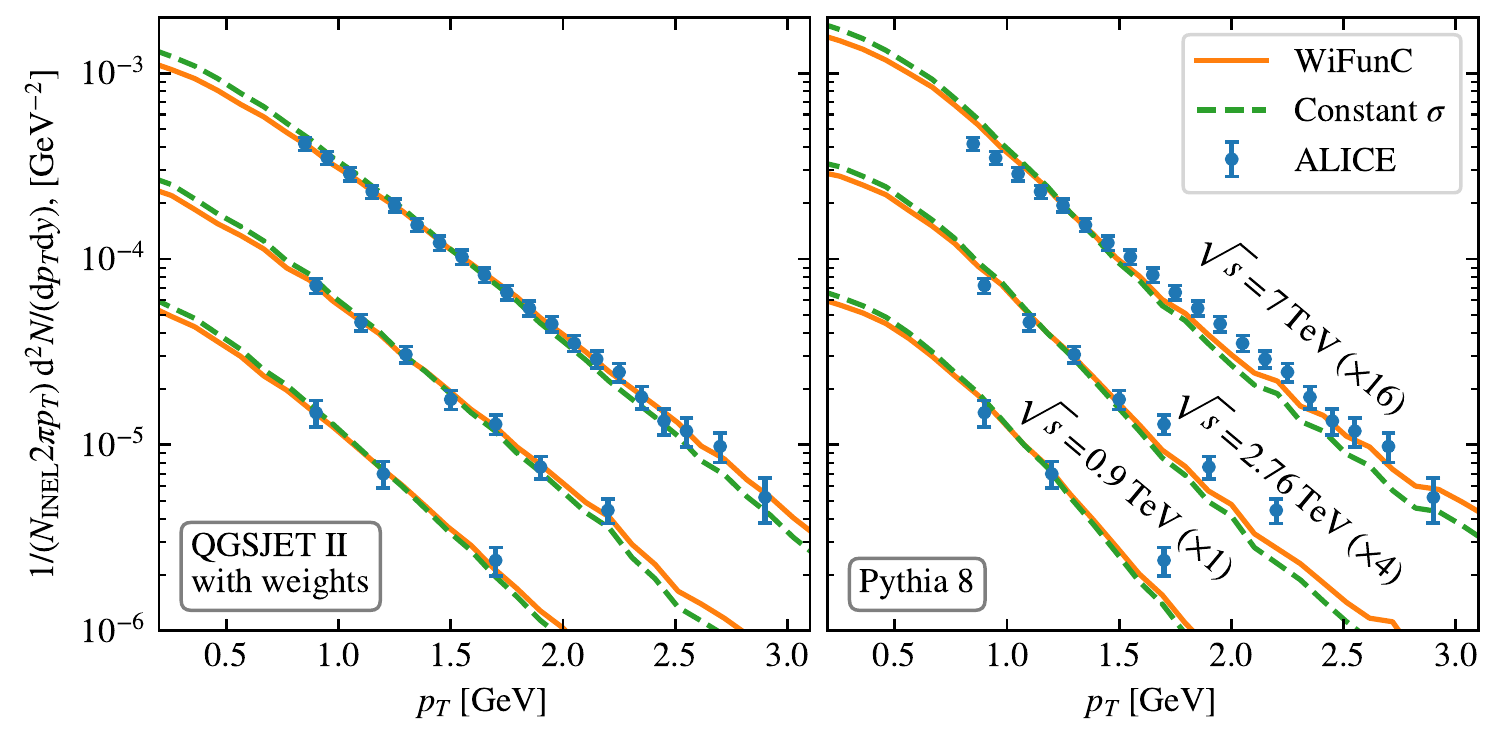}
	\caption{The best fit of the WiFunC model using QGSJET II (left) and
	Pythia 8.2 (right) to the invariant differential yield of antideuterons
	as a function of the transverse momentum measured by the
	ALICE collaboration \cite{collaboration_production_2018} at
	0.9, 2.76 and $7\unit{TeV}$ (blue points).
	The data are multiplied by a constant
	factor to make the figure clearer. In the fit using QGSJET, a weight is 
	included in order to better reproduce corresponding nucleon measurements
	\cite{Kachelriess:2020uoh}. The WiFunC model (orange line) reproduce
	well the data points. The WiFunC model without any Lorentz boost gives a 
	similar fit as the standard coalescence model and is shown for comparison
	(green dashed line).
	}
	\label{fig:alice}
\end{figure}

The WiFunC model improves significantly
the fit to experimental data at large transverse momenta compared to the
standard coalescence model. This is readily 
seen in Fig. \ref{fig:alice}, where the best fit to the invariant differential
yield of antideuterons measured by the ALICE collaboration
\cite{collaboration_production_2018} is plotted using Pythia 8 and QGSJET II. 
Furthermore, it describes the behaviour of the non-trivial baryon emission
volume as measured by the ALICE collaboration
\cite{Kachelriess:2020amp,Acharya:2020dfb}. This is particularly important as
it provides a method of determining the parameter $\sigma$ independent of an
event generator (orange triangle in Fig. \ref{fig:sigma_comparison}).

\subsection{Comment on helium-3 and tritium production}
\label{ssec:he_wf}

In contrast to the deuteron wave function, the helium-3 and tritium wave
functions are not well known. Therefore, the rms charge radii
$r_\mathrm{rms}^{\mathrm{^3He}}=1.96\unit{fm}$ and 
$r_\mathrm{rms}^{t}=1.76\unit{fm}$ were used in Refs.
\cite{Kachelriess:2019taq,Kachelriess:2020uoh} to describe the wave functions
as a Gaussian.\footnote{
	Since the spectra of tritium and helium-3 are expected to be similar, it
	is convenient to assume $r_\mathrm{rms}^{t}=r_\mathrm{rms}^{\mathrm{^3He}}$
	\cite{Kachelriess:2019taq}.
}
This will lead to an artificially small spread
$\sigma$ if in reality either the matter radius is smaller than the charge radius
or the wave function is more peaked than a Gaussian.

In the coalescence picture, deuteron, helium-3 and tritium is expected
to be created by nucleons with the same spread, and the parameter $\sigma$ should
thus be the same for all three particles.
In the previous subsection, we saw that the spread is already well determined by
antideuteron experiments, which has been confirmed by an independent experiment
on the baryon emission volume. Thus, we can allow ourselves to degrade
$r_\mathrm{rms}^{\mathrm{^3He}}$ to a free parameter to better describe the
effective ground state of the nucleus.
In the spirit of simplicity and for consistency
with Eq. \eqref{eq:radius_scaling}, we set
$r_\mathrm{rms}^{^3\mathrm{He}}=b=3^{1/3}\sigma$. Remarkably, this choice 
leads to $\sigma=(0.93\pm0.04)\unit{fm}$ for Pythia 8 and
$\sigma=(1.07\pm0.03)\unit{fm}$ for QGSJET II
when fitted to the antihelium-3 spectrum at
7 TeV measured by the ALICE collaboration \cite{collaboration_production_2018},
consistent with the antideuteron fits in Fig.
\ref{fig:sigma_comparison}. These values 
corresponds to a rms radius 1.3--1.6$\unit{fm}$, which is by no means
far-fetched.\footnote{This change may lead to an increase in the cosmic ray
antihelium spectra in Ref. \cite{Kachelriess:2020uoh} by a factor of a few.
} Nevertheless, it should be emphasised that theoretical
uncertainties related to the wave function has not been properly accounted for.

\subsection{Comment on thermal models}
\label{sec:thermal}

The production of light nuclei in heavy ion collisions is often described using
thermal statistical models, where it is assumed that 
the nuclei are produced around chemical
freeze-out in an expanding ``fireball'' consisting of a quark gluon
plasma. Such models are motivated by the
observation that the nucleus spectra are near thermal with a similar freeze-out
temperature as for nucleons and mesons \cite{Andronic:2017pug}. 
It is, however, hard to resolve
the issue on how particles with small binding energies can survive the 
freeze-out process. Furthermore, since the hadron spectra already are 
thermal-like, coalescence models also predicts thermal-like 
nucleus spectra up to a quantum mechanical correction factor
\cite{Csernai:1986qf}.

Intriguingly, recent observations show characteristic signs for the production
of a quark gluon plasma in small interacting systems, such as $pp$ and $p$Pb
collisions \cite{Nagle:2018nvi}. Such hints have been used as a motivation for
using thermal models to describe the nucleus production even in small
interacting systems \cite{vonDoetinchem:2020vbj,Bellini:2018epz,%
Cleymans:2011pe,Acharya:2020sfy}.
However, many of these, e.g. the behaviour of the $B_2$ factor as a
function of multiplicity and transverse momentum in $pp$ collisions
\cite{Acharya:2020sfy} and the decrease in the baryon emission volume with 
increasing transverse momentum \cite{Acharya:2020dfb}, is also naturally
described by the WiFunC model
\cite{Kachelriess:2020amp}. Moreover, the hints are irrelevant for astrophysical
processes since they are at present only observed at high energies and
multiplicities.

Regardless of the underlying hadron production, the WiFunC model is applicable
as long as the nuclei are produced in a
coalescence process. It is thus worth pointing out that it is argued in Ref.
\cite{Bellini:2020cbj} that two-particle correlation experiments can be used to
validate the coalescence hypothesis, and that the success of the femtoscopy
analysis is strong evidence that coalescence is a major production mechanism for
nuclei.

\section{Numerical implementation of the WiFunC model}
\label{sec:numerical_implementation}

In this section, we briefly summarise how the WiFunC model can be evaluated
for deuteron, helium-3 and tritium. The momentum of the nucleons produced in a
particle collision is to be computed on an event-by-event basis using an
event generator. Double counting can in all relevant applications be
neglected \cite{Kachelriess:2019taq}.

\begin{figure}[htbp]
	\centering
	\includegraphics[scale=.5]{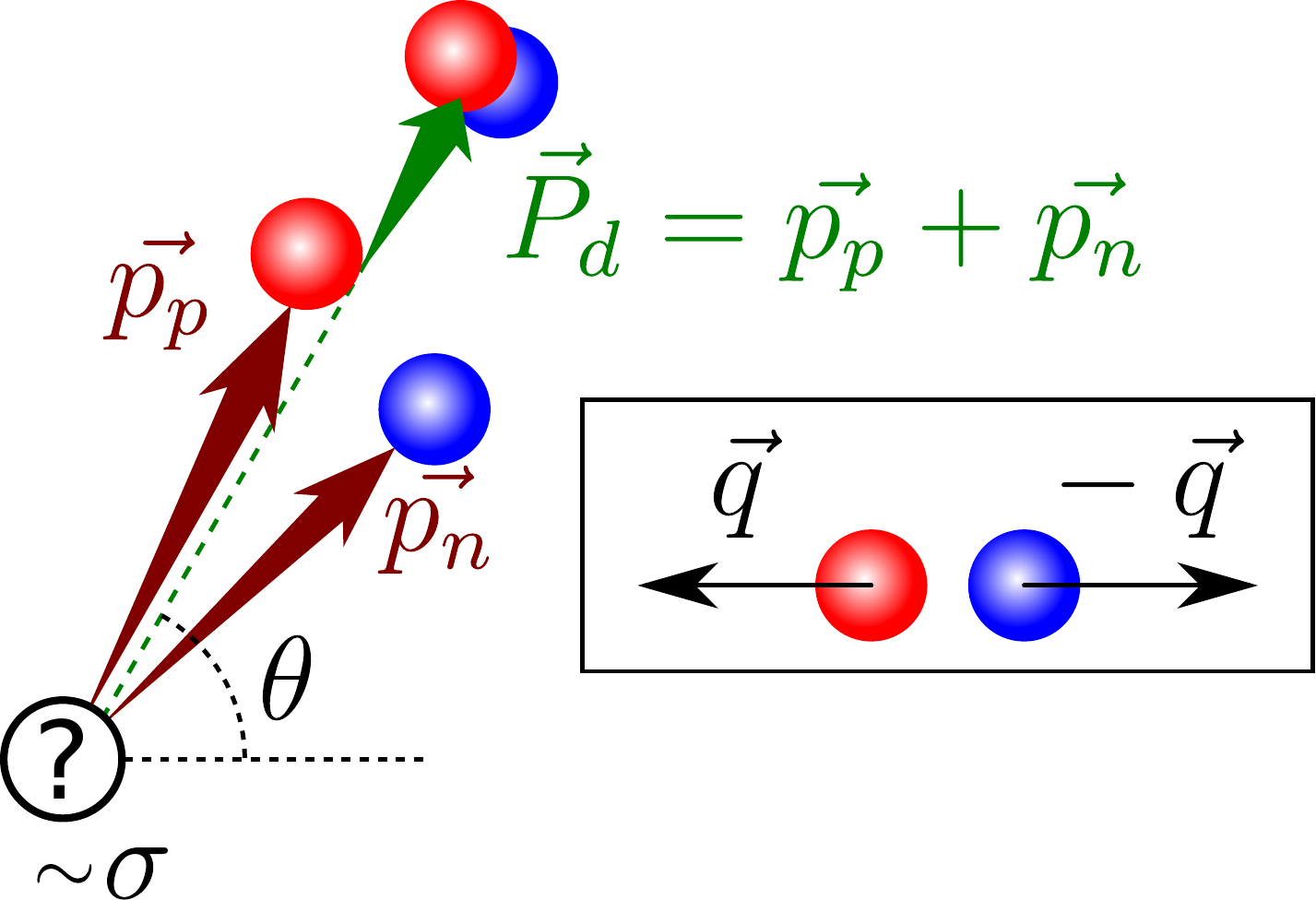}
	\caption{Sketch of the coalescence process in the c.m. frame of the particle
	collision. A proton with momentum $\vec{p}_p$ and neutron with momentum
	$\vec{p}_n$ may coalesce into a deuteron with momentum
	$\vec P_d=\vec{p}_p+\vec{p}_n$ if their momentum difference in the
	deuteron rest frame, $2|\vec{q}|$, is small. The exact probability is
	determined by Eq. \eqref{eq:deuteron_weight}.}
	\label{fig:sketch}
\end{figure}

In the WiFunC model, the probability that a given proton-neutron pair
coalesce is
\begin{equation}
	w = 3\Delta\zeta_1\me^{-d_1^2q^2} + 3(1-\Delta)\zeta_2\me^{-d_2^2q^2},
	\label{eq:deuteron_weight}
\end{equation}
where
\begin{equation}
	\zeta_i = \sqrt{\frac{d_i^2}{d_i^2 + 4\tilde{\sigma}_\perp^2}}
		\sqrt{\frac{d_i^2}{d_i^2 + 4\sigma_\perp^2}}
		\sqrt{\frac{d_i^2}{d_i^2 + 4\sigma_\parallel^2}}, 	
\end{equation}
and $\tilde{\sigma}_\perp^2=\sigma_\perp^2/(\cos^2\theta +\gamma^2\sin^2\theta)$.
The quantities that varies event-by-event are indicated in Fig.
\ref{fig:sketch}: $|\vec{q}|$ is the momentum of the nucleons in the deuteron
rest frame, $\theta$ is the angle between the deuteron momentum and the $z$-axis
of the particle collision in the c.m. frame of the collision and $\gamma$ is the
deuteron Lorentz factor in the c.m. frame of the collision. 
The parameters $\Delta=0.581$, $d_1 = 3.979\unit{fm}$, $d_2=0.890\unit{fm}$ are
fixed by fitting a two-Gaussian wave function to the Hulthen wave function
that well describes the ground state of the deuteron \cite{Kachelriess:2019taq}.
The process dependence of the nucleon spread can be taken into account using Eqs.
\eqref{eq:process_dependence_1}--\eqref{eq:process_dependence_5}.
In particular, the longitudinal and transverse
spreads in $e^+e^-$ and $pp$ collisions can be set equal and is given by
$\sigma\equiv\sigma_{(e^+e^-)} = \sigma_{(pp)}/\sqrt{2}\simeq (1.0\pm0.1)\unit{fm}$.

The model for helium-3 and tritium production is similar to that for deuteron.
In this case, the probability that a proton-proton-neutron or
proton-neutron-neutron triplet coalesce is
\begin{equation}
	w = \frac{64}{12}\zeta^\mathrm{He}\me^{-b^2P^2},
	\label{eq:heluium_weight}
\end{equation}
with
\begin{equation}
	P^2=\frac{1}{3}\left[(\vec p_1-\vec p_2)^2 + (\vec p_1-\vec p_3)^2 +
		(\vec p_2-\vec p_3)^2\right]
	\label{eq:heluium_P2}
\end{equation}
and
\begin{equation}
	\zeta^\mathrm{He} = \frac{b^2}{b^2 + 2\tilde{\sigma}_\perp^2}
		\frac{b^2}{b^2 + 2\sigma_\perp^2}
		\frac{b^2}{b^2 + 2\sigma_\parallel^2}.
\end{equation}
Here, $\vec{p}_i$ ($i=1,2,3$) is the momentum of the nucleons in the rest frame
of the three-particle state. Due to the Gaussian suppression in Eq.
\eqref{eq:heluium_weight}, one can to a good approximation evaluate the
momentum difference of the
nucleons entering Eq. \eqref{eq:heluium_P2} in the corresponding two-particle
rest frames.

In Eq. \eqref{eq:heluium_weight} the nucleus wave function is approximated by 
a Gaussian with a rms radius $b$. In Refs. \cite{Kachelriess:2019taq,%
Kachelriess:2020uoh} it was set equal to the measured helium-3 rms charge radius
$b=1.96\unit{fm}$. However, this choice may lead to an
artificially low nucleus yield. Since $\sigma$ is already well constrained by
antideuteron measurements, $b$ can be degraded to a free parameter e.g. by
setting $b=3^{1/3}\sigma_{(e^+e^-)}$.


\section{Summary and Outlook}
\label{sec:summary}

The WiFunC model is a per-event coalescence model based on the Wigner function
representation of the nucleon and nucleus states. This model includes in a
semi-classical picture both two-nucleon momentum correlations
provided by a Monte Carlo event
generator and the size of the nucleon emission volume.
Since the emission volume is process dependent, it explains naturally the
differences in (anti)nucleus yields observed in $e^+e^-$, $pp$ and $pN$
collisions. The value obtained by fits to experimental data,
$\sigma=(1.0\pm0.1)\unit{fm}$, agrees well with its physical interpretation.

The spread $\sigma$ is currently well constrained by antideuteron data in $pp$
collisions at large energies and large $p_T$. Thus, experimental studies on
antinuclei production in small interacting systems at low energies in the
forward direction is highly warranted. This will reduce uncertainties related
to antinuclei production in astrophysical processes, as well as test the WiFunC
model. Furthermore, two-particle correlation experiments can be used to directly
measure $\sigma$ \cite{Kachelriess:2020amp} and test the coalescence hypothesis
\cite{Blum:2019suo,Bellini:2020cbj}.

The WiFunC model can, in principle, describe the coalescence of
nucleons in any particle collision as long as the event generator is able
to describe the underlying physics. Currently, the prescription of $\sigma$
is expected to describe well coalescence is small interacting systems, such as
$e^+e^-$, $pp$ and $pN$ collisions. However, the
uncertainties in $\sigma$ will in general increase with increasing
system size. Therefore,
it would be interesting to instead use the space-time descriptions implemented
in event generators like Pythia
\cite{sjostrand_pythia_2006,sjostrand_introduction_2015,Ferreres-Sole:2018vgo}
or UrQMD \cite{Bleicher:1999xi,Sombun:2018yqh} by
evaluating Eq. \eqref{eq:wifunc_with_r} directly.

\acknowledgments
I would like to thank S. Ostapchenko and M. Kachelrie\ss~ for fruitful
collaborations on which this review is based, M. Kachelrie\ss~ for
comments on this text.

\providecommand{\href}[2]{#2}\begingroup\raggedright\endgroup

\end{document}